\documentclass[]{emulateapj}
\usepackage{graphicx}
\usepackage{latexsym}
\usepackage{natbib}
\begin{document}

\title{Millihertz Oscillation Frequency Drift Predicts the Occurrence
of Type I X-ray Bursts}

\author{D. Altamirano\altaffilmark{1}, 
M. van der Klis\altaffilmark{1}, 
R. Wijnands\altaffilmark{1} 
A. Cumming\altaffilmark{2}}

\email{diego@science.uva.nl}

\altaffiltext{1}{Astronomical Institute, ``Anton Pannekoek'',
University of Amsterdam, and Center for High Energy Astrophysics,
Kruislaan 403, 1098 SJ Amsterdam, The Netherlands.}

\altaffiltext{2}{Department of Physics, McGill University,
3600 rue University, Montreal, QC, H3A 2T8, Canada}

\begin{abstract}

Millihertz quasi-periodic oscillations reported in three neutron-star
low mass X-ray binaries have been suggested to be a mode of marginally
stable nuclear burning on the neutron star surface. In this Letter, we
show that close to the transition between the island and the banana
state, 4U~1636--53 shows mHz QPOs whose frequency systematically
decreases with time until the oscillations disappear and a Type I X-ray
burst occurs.
There is a strong correlation between the QPO frequency $\nu$ and the
occurrence of X-ray bursts: when $\nu\gtrsim9$~mHz no bursts occur,
while $\nu\lesssim9$~mHz does allow the occurrence of bursts.
The mHz QPO frequency constitutes the first identified observable that
can be used to predict the occurrence of X-ray bursts.  If a
systematic frequency drift occurs, then a burst happens within a few
kilo-seconds after $\nu$ drops below 9~mHz. This observational result
confirms that the mHz QPO phenomenon is intimately related with the
processes that lead to a thermonuclear burst.

\end{abstract}
\date{\today}
\maketitle

\section{Introduction}
\label{sec:intro}

\citet{Revnivtsev01} discovered a new class of low frequency
quasi-periodic oscillation (QPO) in three neutron star X-ray binary
sources (Aql X-1, 4U~1608--52 and 4U~1636--53). This new QPO has
frequencies between 7 and 9 $\times 10^{-3}$ Hz, i.e. it is in the
milli-Hertz range, and its other properties also differ from those of
the other QPOs found in the neutron star systems
\citep[e.g. energy dependence, see][]{Vanderklis06}.
 Although \citet{Revnivtsev01} could not discard an interpretation
related with  disk instabilities, they  concluded that
the mHz QPO is likely due to a special mode of nuclear
burning on the neutron star surface. 
This interpretation was strengthened by the results of \citet{Yu02},
who showed that the kHz QPO frequency is anti-correlated with the
luminosity variations during the mHz oscillation, suggesting that the
inner edge of the disk slightly moves outward as the luminosity
increases during each mHz cycle due to stresses generated by radiation
coming from the the neutron star surface.
This is contrary to the correlation observed between X-ray luminosity
($L_x$) and kHz QPO frequency, where the inner disk edge is thought to
move in, as the accretion rate and hence $L_x$, increases \citep[][and
references within]{Vanderklis06}.

The properties of the mHz QPOs as observed up to now can be summarized
as follows:
\textit{(1)} the fractional rms amplitude strongly decreases with
energy, from $\approx2$\% at 2.5 keV, down to an almost undetectable $ <
0.2$\% at $\approx5$ keV;
\textit{(2)} mHz QPOs occur only in a particular range of  X-ray
luminosity: $L_{2-20~keV} \approx 5-11 \times 10^{36}$ erg/s;
\textit{(3)} the frequency of the mHz QPOs is  between 7 and
9 mHz;
\textit{(4)} the mHz QPOs disappear with the occurrence of a type I X-ray burst;
\textit{(5)} as noted above, the kHz QPO frequency is approximately
anti-correlated with the $2-5$ keV count rate variations that constitute the 
mHz oscillation.

\citet{Heger07} suggested that the mHz QPOs could be explained as the
consequence of marginally stable nuclear burning on the neutron star
surface.  They found an oscillatory mode of burning, with a period
$P_{osc}$ close to the geometric mean of the 
thermal\footnote{The thermal timescale is defined as $t_{thermal}=c_p
T / \epsilon$ where $c_p$, $T$ and $\epsilon$ are the heat capacity at
constant pressure , the temperature and the nuclear energy generation
rate, respectively}
and accretion\footnote{The accretion timescale is defined as
 $t_{accr}=y /\dot{m}$ where $y$ and $\dot{m}$ are the column depth
 of the burning layer and local accretion rate, respectively.}
timescales of the burning layer. For typical parameters, 
$P_{osc}\equiv\sqrt{t_{thermal} \cdot t_{accr}}\approx2$ minutes, in
accordance with the characteristic frequency of the mHz QPOs.  The
burning is oscillatory only close to the boundary between stable
burning and unstable burning (in Type I X-ray bursts), explaining the
observation that the mHz QPOs were seen within a narrow range of
luminosities.

Two of the three sources in which \citet{Revnivtsev01} found the mHz
QPOs are transient atoll sources (Aql X-1 and 4U~1608--52) while the
third one, 4U~1636--53, is a persistent atoll source
\citep{Hasinger89}. The object of our current study, 4U~1636--53, has
an orbital period of $\approx3.8$ hours \citep{Vanparadijs90} and a
companion star with a mass of $\approx0.4 \ M_{\odot}$ \citep[assuming a
NS of 1.4~$M_{\odot}$; ][]{Giles02}.  4U~1636--53 is an X-ray burst
source \citep{Hoffman77} showing asymptotic burst oscillation
frequencies of $\approx581$~Hz \citep[][]{Zhang97,Strohmayer02}.  The
aperiodic timing behavior of 4U~1636--53 has been studied with the
EXOSAT Medium Energy instrument \citep[][]{Prins97} and with the Rossi
X-ray Timing Explorer  \citep[RXTE, e.g.][]{Wijnands97,Disalvo03,Altamirano07}.
%%%%%%%%%%%%%%%%%%%%%%

\begin{figure}[t] 
\center
\resizebox{1\columnwidth}{!}{\rotatebox{0}{\includegraphics{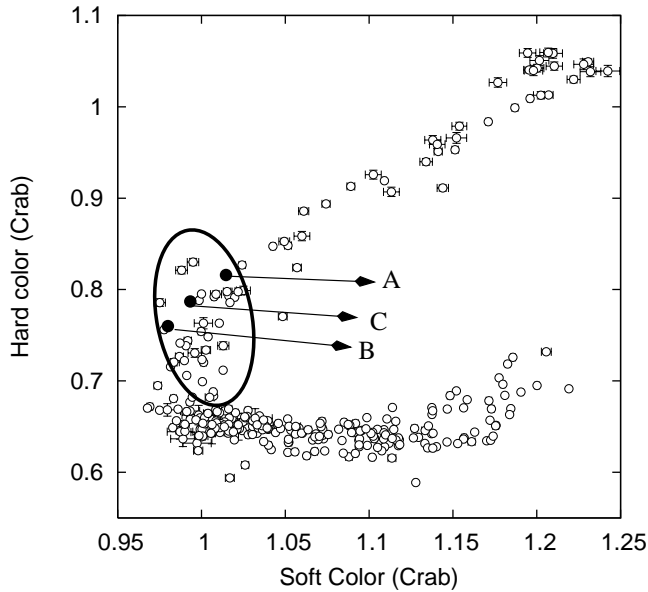}}}
\caption{Color-color diagram as \citet{Altamirano07}. Each data point
represents the average of an observation ($\approx2$ to
$\approx30$~ksec). The ellipse marks the region in which mHz QPOs with
decreasing frequency were found. The labels A, B and C correspond to
those in Figure~\ref{fig:lc}.
}
\label{fig:ccd}
\end{figure}

\begin{figure}[t] 
\center
\resizebox{1\columnwidth}{!}{\rotatebox{0}{\includegraphics{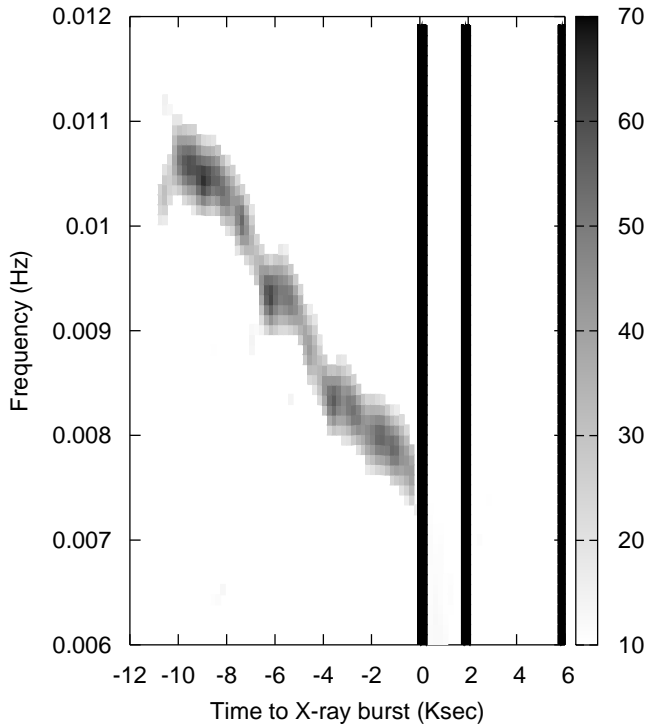}}}
\caption{Dynamical power spectrum smoothed with a 750 seconds sliding
window with steps of 200 seconds, showing the mHz QPOs during the last
12~ksec before the X-ray burst occurs. This sequence corresponds to
case B in Figures~\ref{fig:lc}~\&~\ref{fig:ccd} -- ObsId:
60032-05-02-00. The three black vertical lines correspond to the times
of occurrence of the X-ray bursts. For clarity, we plot only powers
above 10 which correspond to $\gtrsim2\sigma$ (single trial per 750
seconds window but normalized to the number of possible frequencies in
the range 0--0.5~Hz).}
\label{fig:dynspec}
\end{figure}

\begin{figure*}[t] 
\center
\resizebox{2\columnwidth}{!}{\rotatebox{0}{\includegraphics{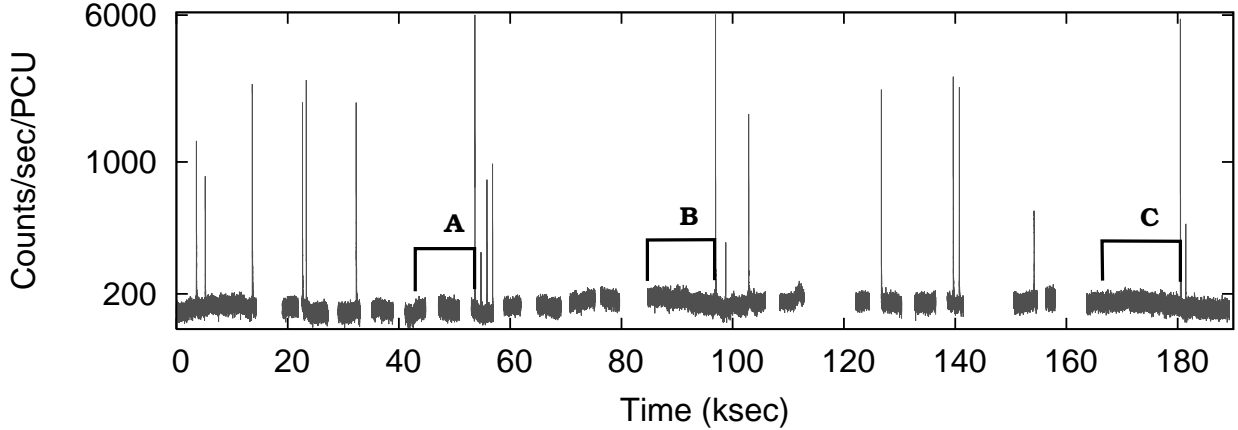}}}
\caption{2--60~keV PCA light curve of observations
60032-05-01,02,03 and 04. $A$, $B$ and $C$ indicate intervals in which the
mHz QPOs are detected. In all three cases, the frequency of the mHz
QPO decreases with time until the appearance of an X-ray burst. }
\label{fig:lc}
\end{figure*}

4U~1636-53 is a reference source for studying nuclear burning on the
 surface of a neutron star since it shows the full range of burst
 behavior: single and multi-peaked Type I X-ray bursts, superbursts,
 burst oscillations, photospheric radius expansion, regular and
 irregular burst sequences \citep[e.g.][]{Galloway06} and mHz QPOs. As
 such, it is an ideal source to understand the relation between these
 different observational manifestations of nuclear burning.
 
Recently, \citet{Shih05} reported that 4U~1636--53 has shown a
significant decrease in its persistent $L_x$ during the years 2000 and
2001. \citet{Altamirano07} show that during the low $L_x$ period,
4U~1636--53 is observed in its (hard) island states. This provides an
opportunity to study the mHz QPOs in harder and lower luminosity
states than was possible up to now.

\section{DATA ANALYSIS \& RESULTS}
\label{sec:dataanalysis}

We used data from the RXTE Proportional Counter Array and the High
Energy X-ray Timing Experiment \citep[PCA and HEXTE, respectively; for
instrument information see][]{Jahoda06,Gruber96}. Up to June, 2006,
there were 338 public pointed observations. An observation covers 1 to
5 consecutive 90-min satellite orbits. Usually, an orbit contains
between 1 and 5 ksec of useful data separated by 1--4 ksec data gaps;
on rare occasions the visibility windows were such that RXTE
continuously observed the source for up to 27~ksec. In total there
were 649 gap-free data segments of length 0.3 to 27 ksec.

We produced energy spectra for each observation using Standard data
 modes and fitted them in the $2-25$~keV and $20-150$~keV bands for
 PCA and HEXTE, respectively. The interstellar absorption $N_H$ was
 fixed at $3.75\times10^{21}$ cm$^{-2}$
 \citep[see][]{Schulz99,Fiocchi06}.  We used 1-sec resolution event
 mode PCA light curves in the $\approx2-5$~keV range (where the mHz QPOs
 are strongest) and searched for periodicities in each of the 649
 segments separately using Lomb-Scargle periodograms
 \citep{Lomb76,Scargle82,Numerical-Recipes}. Segments in which one or
 more Type I X-ray bursts were detected were searched for
 periodicities before, in between, and after the bursts.
We find that the oscillations in the $\approx2-5$~keV range are
evident from the light curves \citep[see for example figure 1
in][]{Revnivtsev01}. The significance as estimated from our
Lomb-Scargle periodograms \citep[][]{Numerical-Recipes} confirm that
the oscillations are all above the $3\sigma$ level. We estimated the
uncertainties in the measured frequencies by fitting a sinusoid to
1000~sec data segments to minimize frequency-drift effects.  The
typical errors on the frequency are of the order of $2-6 \times
10^{-5}$~Hz (or $2-6 \times 10^{-2}$~mHz). 

We detected mHz QPOs in 124 of the 649 segments. Most occur in
segments with less than 4~ksec of useful data and sometimes the QPOs
cover only part of a segment.
\citet{Revnivtsev01} reported the characteristics of the mHz QPOs
 between March 1996 and February 1999. Using the X-ray colors averaged
 per observation as reported by \citet{Altamirano07}, we find that
 their data sample the region at hard colors $\lesssim0.7$ and soft
 colors $\gtrsim1$ (see Figure~\ref{fig:ccd}), which represents the so
 called banana state \citep{Vanderklis06}.
Some of the later observations also sample the banana state. We
re-analyzed all the data in this region of the color-color diagram and
found results which are consistent with those reported by
\citet{Revnivtsev01}: the frequency of the QPOs varies randomly
between $6$ and $9$~mHz.

In the harder state close to the transition between the island and
banana state and marked with the ellipse drawn in Figure~\ref{fig:ccd},
we found 22 segments with significant mHz QPOs;
in these observations, the (2--150~keV) luminosity was
$6-10\times10^{36} (d/(6 kpc))^2 \ $erg s$^{-1}$, while for the other
cases, corresponding to the banana state, it was higher
($10-35\times10^{36} (d/(6 kpc))^2 \ $erg s$^{-1}$).

Among the 22 segments, we distinguish two groups based on segment
length: the first consisting of 4 segments each with more than 14
ksec of uninterrupted data, and the second consisting of 18 segments
each corresponding to one orbit with less than $\approx4$~ksec of useful
data.
For all four segments in the first group, we measure a systematic
decrease of frequency from between $10.7$ and $12.5$~mHz down to less
than $9$~mHz over a time interval of 8 to 12 ksec, after which an
X-ray burst occurs and the QPOs disappear (the QPOs become
much less than $3\sigma$ significant).
Figure~\ref{fig:dynspec} shows a representative dynamical power spectrum
corresponding to one of these segments (interval B in Figure~\ref{fig:lc}).
The QPO is present $\approx12$~ksec seconds before the burst and its
frequency systematically decreases with time from $\approx10.7$~mHz down
to $\approx7.6$~mHz. Then the X-ray burst occurs and the QPO disappears.
In the second group, 16 of the 18 segments of $<4$~ksec show the mHz
QPO frequency to decrease either within a segment, or between 2 or 3
consecutive orbits (with 2--4 ksec data gaps in between) at rates
consistent with those seen in the 4 long segments.
The two remaining segments are too short and isolated to constrain
the frequency drift well.

To illustrate the interplay between this very systematic behavior of
the mHz QPOs and our data structure, in Figure~\ref{fig:lc} we show a
representative lightcurve.
A, B and C mark three intervals in which the mHz QPOs were detected
and that each terminate with an X-ray burst. As can be seen, we have
data in which mHz QPOs are detected and followed through consecutive
segments (case A), and data in which the oscillations are detected and
disappear within one segment (cases B \& C). Furthermore, we have data
in which the oscillations are present from the start of the
observation (case B) as well as data in which the mHz QPOs appear
during an observation (case A \& C).

Among our 22 segments, the frequency of the oscillations
varies in the range $7-14.3$~mHz with directly observed onset
frequencies between $10.7$ and $14.3$~mHz.
Interpolating through gaps, the QPOs last for 7.5 to 16
ksec. Over such intervals, the frequency is always consistent
with decreasing at average rates from $0.07$ to $0.15$ mHz
ksec$^{-1}$, and
 the frequency had always dropped to $\lesssim9$~mHz just before an
X-ray burst (as estimated from the last 750 seconds before the
burst). Interestingly, this last result applies to all cases in which
we detect the mHz QPOs before an X-ray burst, including the cases that
occur in the banana state: it seems that independent of the spectral
state of the source, no X-ray burst will occur if the mHz QPOs are
present at a frequency higher than $\approx9$~mHz (bursts do occur in
both states that are not preceded by detectable mHz QPOs).

No relation between the 2--60~keV count rate and frequency was found:
in two of the four long segments the count rate decreased about 10\%
during the time the mHz QPOs were present, while in the other two
cases the count rate increased by approximately the same amount. No
clear relation was found between frequency range covered and duration
of the oscillation; perhaps this is related to the fact that, as shown
in Figure~\ref{fig:dynspec}, the frequency does not decrease smoothly
but has short periods in which it is consistent with being
constant.

When 4U~1636--53 is observed close to its island--banana state
transition, the mHz QPOs disappear \textit{only} when an X-ray burst
occurs. However, this is not the case for the banana state, in which
we found also observations in which the oscillation disappear below
detectable levels without the occurrence of an X-ray burst.
The interval of time required to
again detect the oscillations after a burst occurred is variable.
The two extreme cases are
(i) observation 60032-01-06-000 where no mHz QPOs were detected
during the $\approx15$~ksec of uninterrupted data following an X-ray
burst and
(ii) observation 40028-01-06-00, where mHz QPOs are detected again
$\approx6000$ seconds after an X-ray burst occurred.
We note that in the first case, the source was close to the transition
between island and banana state while in the second case, the source
was in the banana state. Nevertheless, no clear relation between this
waiting time and the source state (island or banana state) was found.
As bursts may be missed due to data gaps, a time interval of
$\approx1000$ seconds between a (missed) X-ray burst and onset of the
QPOs cannot in some cases be excluded (e.g. case A in
Figure~\ref{fig:lc}).

\section{Discussion}\label{sec:discussion}

We have shown that close to the transition between the island and the
banana state 4U~1636--53 exhibits mHz QPOs whose frequency
systematically decreases with time until the oscillations disappear
with the occurrence of a Type I X-ray burst. The mHz QPO frequency
$\nu$ constitutes the first identified observable that can be used to
predict the occurrence of X-ray bursts: when $\nu\gtrsim9$~mHz no
bursts occur, while $\nu\lesssim9$~mHz does allow the occurrence of
bursts. If a systematic frequency drift occurs, then a burst happens
within a few kilo-seconds after $\nu$ drops below 9~mHz. This
observational result confirms that the mHz QPO phenomenon is
intimately related with the processes that lead to a thermonuclear
burst.

The fact that the observation of a systematic frequency decrease with
time implies the occurrence of a future X-ray burst, strongly suggests
that the frequency of the mHz QPOs is related to the burning processes
on the neutron star surface. One possibility is that the frequency of
the QPO is somehow a measurement of the accumulation of fresh fuel on
the neutron star surface which will be available for a future
thermonuclear burst.
To our knowledge, there has been only one attempt to theoretically
explain the mHz QPOs phenomena \citep{Heger07}. In this model the
frequency of the QPO depends, among others, on the amount of available
fresh fuel, on the local accretion rate and the composition of the
material. It is beyond the scope of this Letter to perform 
numerical simulations as those reported by \citet{Heger07}.
In the rest of this discussion we briefly compare these authors' model
predictions with our observations and propose some more complex
scenarios.

Analytical and numerical results based on the simplified one-zone
model of \citet{Paczynski83} in the \citet{Heger07} marginally stable
burning model (see Section~\ref{sec:intro}) predict that 
(i) close to the boundary between stable and unstable burning, the NS
surface will show temperature fluctuations with constant frequency
$\nu$ if the local accretion rate $\dot{m}$ remains constant;
(ii) this marginally stable burning regime will occur at $\dot{m}$ near
Eddington, hence accretion must be confined to a surface area $S_A$
that is much smaller that the total area of the NS;
(iii)  $\nu$ correlates with $\dot{m}$ \citep[see figure 4
in ][]{Heger07} and
(iv) thermonuclear bursts and mHz QPOs should not be observed at the same
luminosity and therefore presumably at the same $\dot{m}$.

In this paper, we show that for constant luminosity the QPO frequency
can systematically decrease in time
and that instantaneously measured frequencies can be the same for
 different luminosities.
% and that when no frequency drift is observed,
% the frequency measured is consistent with being the same for
% different \textbf{banana state} luminosities. 
We also show that mHz QPOs and thermonuclear bursts do in fact occur
 at the same luminosity and that both phenomena are clearly related.
This means that we are dealing with a more complex scenario than that
introduced by \citet{Heger07}.

The amount of time between the preceding X-ray burst and the onset of
mHz QPOs is variable ($>6$ksec) and apparently independent of source
state.
If the system is locally accreting at $\dot{m}\simeq\dot{m}_{Edd}$
%\simeq8\cdot10^4$ g~cm$^{-2}$~s$^{-1}$ 
and if none of the accreting fuel is burnt, only $\approx1000$~seconds
are required to accrete a fuel layer of column depth $y_f$
capable of undergoing marginally stable burning
\citep[$y_f\approx10^8$~g~cm$^{-2}$ and $\dot{m}\approx8\cdot10^4$
g~cm$^{-2}$~s$^{-1}$ -- see e.g.][]{Heger07}. One possible explanation
for the observed longer intervals between burst and onset of
oscillations, is that a large fraction of the accreted fuel is burnt
as it is accreted on the neutron star surface.
Of course the burning fraction could vary in time, and this estimate
is assuming that all the fuel was burnt during the last X-ray burst,
which is not always true \citep{Bildsten98}. 
Interestingly, if this interpretation is correct and low partial
burning fractions can occur, under certain conditions the mHz QPOs
could appear in much less than a 1000 seconds after an X-ray
burst.

The fact that the amount of time between the preceding X-ray burst and
the onset of mHz QPOs is variable may be also an indication that not
all the accreted fuel is burnt nor available to participate in the
marginally stable burning.
For example, accretion could occur onto an equatorial region occupying
less than 10\% of the surface area of the star \citep{Heger07}. A
possibility is that part of the fresh fuel burns stably at a rate
$B(t)$ per unit area while the other part leaks away from this region
at a rate $R(t)$. While the material accumulated at a rate $R(t)$
would serve as fuel for a thermonuclear burst, marginally stably
burning of the matter on the equatorial belt is (in principle) still
possible.
Although such scenario cannot explain the frequency drifts we observe,
it can explain why mHz QPO and X-ray bursts do occur at the same $\dot
m$. If mHz QPOs can only occur at a certain local accretion rate
$\dot{m}$ ($\simeq\dot{m}_{Edd}$), a small change in effective local
accretion rate will lead to an absence of mHz QPOs. This might explain
why the mHz QPOs are not always present between X-ray bursts.

Another possibility (which is not taken into account in
\citet{Heger07}'s model) is that there is a significant heat flux from
deeper in the star that heats the region undergoing marginally stable
burning. 
For example, changes in heat flux due to energy that is first
conducted into deeper layers during an X-ray burst and then slowly
outwards towards the surface might be possible.
Such a change in the heat flux could affect the conditions of the
burning layer (e.g. temperature or burning rate $B(t)$) and therefore
could affect the characteristics of the burning processes on the
neutron star surface.

Other aspects of the observations offer further challenges for
theoretical models that explain burning processes on the neutron star
surface as well those which explain atoll sources states.
In particular, 
(i) why the systematic frequency drifts are observed close to the
transition between the island and the banana state while the
frequencies are approximately constant in the banana state. This may
be another indication that the disk geometry of the system is changing during
the state transition \citep[see e.g. ][]{Gierlinski02};
(ii) why in the transition between island and banana state the
oscillations disappear \textit{only} when an X-ray burst occurs, while
in the banana state they can also disappear without an X-ray burst
(see Section~\ref{sec:dataanalysis}).
Clearly, further theoretical work is needed. More observational work
on the interactions between mHz QPOs and X-ray bursts is in progress
and will provide further clues for theoretical models.

{ \footnotesize 
{ 
 \hspace{0.03cm} \textbf{Acknowledgments:} DA wants to thank
  A. Patruno, P. Casella, P. Uttley, M. Linares for very helpful
  discussions. This work was supported by the ``Nederlandse
  Onderzoekschool Voor Astronomie'' (NOVA), i.e., the ``Netherlands
  Research School for Astronomy'', and it has made use of data
  obtained through the High Energy Astrophysics Science Archive
  Research Center Online Service, provided by the NASA/Goddard Space
  Flight Center. AC is grateful for support from NSERC, Le Fonds
  Qu\'eb\'ecois de la Recherche sur la Nature et les Technologies, the
  Canadian Institute for Advanced Research, and as an Alfred P.~Sloan
  Research Fellow.}  }

%\clearpage
%\bibliographystyle{aa} 
%\bibliography{biblio}

\end{document}